\documentclass[english,twocolumn,showpacs,preprintnumbers,prb,amsmath,amssymb]{revtex4}
\usepackage[T1]{fontenc}
\usepackage[latin1]{inputenc}
\usepackage{graphicx}
\usepackage{amssymb}

\makeatletter

\providecommand{\tabularnewline}{\\}

\makeatletter

\makeatletter

\makeatletter

\makeatletter

\usepackage{dcolumn}

\usepackage{bm}

\makeatother

\makeatother

\makeatother

\makeatother

\usepackage{babel}
\makeatother
\begin{document}

\title{Preferential affinity of calcium ions to charged phosphatidic-acid
surface from a mixed calcium/barium solution: X-ray reflectivity and
fluorescence studies}

\author{Wei Bu, Kevin Flores, Jacob Pleasants, and David Vaknin\footnote{vaknin@ameslab.gov
}}

\affiliation{Ames Laboratory, and Department of Physics and Astronomy, Iowa State
University, Ames, Iowa 50011, USA}
\date{\today}
\begin{abstract}
X-ray reflectivity and fluorescence near total reflection experiments
were performed to examine the affinities of divalent ions ($\mathrm{Ca^{2+}}$
and $\mathrm{Ba^{2+}}$) from aqueous solution to a charged phosphatidic-acid
(PA) surface. A phospholipid (1,2-Dimyristoyl-sn-Glycero-3-Phosphate,
DMPA), spread as a monolayer at the air/water interface, was used
to form and control the charge density at the interface. We find that
for solutions of the pure salts (i.e., $\mathrm{CaCl_{2}}$ and $\mathrm{BaCl_{2}}$),
the number of bound ions per DMPA at the interface is saturated at
concentrations that exceed $\mathrm{10^{-3}M}$. For a 1:1 $\mathrm{Ca^{2+}/Ba^{2+}}$
mixed solutions, we find that the bound $\mathrm{Ca^{2+}/Ba^{2+}}$
ratio at the interface is 4:1. If the only property determining charge
accumulation near PA were the ionic charges, the concentration of
mixed $\mathrm{Ca^{2+}/Ba^{2+}}$ at the interface would equal that
of the bulk. Our results show a clear specific affinity of PA for Ca compared to Ba. We provide
some discussion on this issues as well as some implications for biological
systems.
Although our results indicate an excess of counterion charge with respect to the surface charge, that is, charge inversion, the analysis of both reflectivity and fluorescence do not reveal excess of co-ions (namely, $\mathrm{Cl^{-}}$ or $\mathrm{I}^{-}$).


\end{abstract}
\maketitle


\section{Introduction}
Varying affinities among same-charge ions to counterions are expected
and are commonly accounted for in chemistry. Although
they are also common in biological processes, they are less pronounced
because the strong tendency of the aqueous environment in biological
systems to accentuate the electrostatic interactions rather than the
chemical ones. However, close to charged interfaces, in channel proteins,
or at high salt concentrations, the electrostatic description fails
at the relevant short distances and quantum effects set in.
We have undertaken the present study to determine the preferential
affinity of $\mathrm{Ca^{2+}}$ or $\mathrm{Ba^{2+}}$ to a highly
charged interface formed by phosphatidic acid (PA) using X-ray reflectivity and spectroscopy techniques.
While PA is a minority lipid in cell membranes, it participates in numerous biological processes that are commonly triggered by small changes in Ca concentration.\cite{Meijer2003,Ishii2004,Wang2006,Kooijman2007} Molecular dynamics simulations have predicted that PA has a particular affinity to divalent ions via its oxygen atoms that may lead to charge inversion (namely, the charge density of the bound
counterion exceeds that required to neutralize the surface charge
exerted by PA).\cite{Faraudo2007b} These simulations also predict that the PA becomes doubly
charged by proton release.  Recent X-ray experiments on PA monolayers have revealed that charge accumulation of divalent ions and trivalent ions\cite{Vaknin2003,Pittler2006} near PA interfaces show charge inversion, i.e. accumulation of charges that exceed the nominal surface charge. It is therefore important for both physical chemistry and biology to understand how calcium and other divalent ions interact with PA.

X-ray scattering techniques to determine ion distributions and binding
to Langmuir monolayers\cite{Bloch1985,Bloch1988,Kjaer1989,Jun1990,Daillant1991,Jacquemain1991,Novikova2003,Zheludeva2003,Vaknin2003,Bu2005,Bu2006,Bu2006b,Pittler2006,Shapovalov2006,Shapovalov2007}
(LM) have been extensively used to address the long standing problem
of the structure of the electric double layer.\cite{Ninham1971,Bloch1990,Andelman1995,Israelachvili2000}
These include X-ray reflectivity\cite{Kjaer1989,Daillant1991} and
anomalous reflectivity,\cite{Vaknin2003,Bu2005,Bu2006,Pittler2006}
which provide the total number of ions and the spatial ionic distributions
at the charged interface. The near total X-ray reflection fluorescence
(NTRF) technique monitors specifically and quantitatively ions near
the interface\cite{Bloch1985,Bloch1988,Daillant1991,Zheludeva2003,Shapovalov2006,Shapovalov2007}
but does not yield spatial information on the counterion layer. The
near resonance X-ray energy scans at fixed momentum transfers can
probe specifically the interfacial ions and yield their distributions
as well as the $f^{\prime}(E)$ and $f^{\prime\prime}(E)$ fine structures\cite{Bu2006b}
which, like the extended X-ray absorption fine structure technique
(EXAFS), gives information on the local environment of the probed
ion.

While monovalent ion distributions do not show appreciable specificity,
as evidenced by the fact that their behavior are well described by
PB theory that accounts for proton transfer and release \cite{Ninham1971,Bloch1990,Israelachvili2000}
from the charged headgroup,\cite{Bu2005,Bu2006,Bu2006b,Giewekemeyer2007}
distributions for multivalent ions show a high degree of ionic specificity.\cite{Koelsch2007}
The theoretical understanding of such ionic specificity is not fully understood. From classical statistical mechanics, it can be shown that Bjerrum pairing can account for some specificity related to the different sizes of the ions\cite{Travesset2006,Faraudo2007a}, but the general trends observed in many experiments suggest a much stronger specificity, thus requiring more complex quantum chemistry calculations.

Recently Shapovalov and coworkers\cite{Shapovalov2007} applied X-ray
fluorescence studies below the critical angle for total reflection
to behenylsulfate monolayers that are spread on a variety of mixed
ionic solutions to determine preferential association of ions to the
highly charged sulfate head group. In their study of a 1:1 $\mathrm{Ca^{2+}/Ba^{2+}}$
mixed solution, they find that $\mathrm{Ba^{2+}}$ cations have roughly
10-fold preference over $\mathrm{Ca^{2+}}$ to associate with the
sulfate headgroup. Although these authors argue that this result cannot
be explained by the small difference in the hydration-sphere radii,
recent theoretical work suggests the opposite\cite{Biesheuvel2007}.

\section{Experimental Setup and Methods}
Ion bulk concentrations were prepared using solutions of calcium chloride
and iodide ($\mathrm{CaCl_{2}},\mathrm{CaI_{2}}$) and barium chloride
($\mathrm{BaCl_{2}}$) obtained from Sigma-Aldrich. Each solution
was mixed into ultra-pure water (Millipore apparatus; resistivity
18.1 M$\Omega$cm). The surface charge density was controlled by a
monolayer of 1,2-Dimyristoyl-sn-Glycero-3-Phosphate (Monosodium Salt,
DMPA; MW = 614, from Avanti Lipids, Inc.). Small amounts (mili-grams)
of DMPA were mixed into a 3:1 chloroform/methanol solution and spread
at the surface of the bulk concentrations in a temperature controlled
Langmuir trough. The trough is placed in a sealed canister that is
kept at a constant temperature ($22^{o}C$) and is continuously purged
with water-saturated helium gas. 10-15 minutes were allotted for solvent
evaporation after which the monolayer was compressed at a rate of
$1$ {\AA}$^{2}$ per molecule per minute to the desired pressure
(the experiments reported here were conducted at a constant pressure
$\pi\approx$ 25mN/m). During the monolayer compression the surface
pressure was recorded by a micro-balance using a filter-paper Wilhelmy
plate.\cite{Vaknin2003}

X-ray reflectivity and fluorescence studies of monolayers at gas/water
interfaces were conducted on a home built liquid surface X-ray reflectometer,
using an Ultra-X18 Rigaku X-ray source generator with a copper rotating
anode operating at $\sim15$ kW power. Down stream from the beam source,
the characteristic Cu $\mathrm{K\alpha}$ ($\lambda=1.54$ \AA) is
selected and deflected onto the liquid surface at a desired angle
by a Ge(111) crystal. Scattered photon intensities were normalized
to an incident beam monitor placed in front of the sample. The X-ray
diffractometer was used in two configurations: reflectivity and fluorescence.

Specular X-ray reflectivity experiments yield the electron density
(ED) profile across the interface, and can be related to molecular
arrangements in the film.\cite{Vaknin1991,Kjaer1994,Vaknin2003b}
The reflectivity for a LM at $Q_{z}$ is calculated by \begin{equation}
R(Q_{z})=R_{0}(Q_{z})e^{-Q_{z}^{2}\sigma^{2}},\end{equation}
 where $R_{0}(Q_{z})$ is the reflectivity due to an ED profile composed
of step functions, calculated by the recursive dynamical method,\cite{Parrat1954}
and $\sigma$ is an effective surface roughness, which has a weak
$Q_{z}$ dependence according to capillary wave theory applied to
X-ray reflectivity studies.\cite{Braslau1988,Ocko1994} In our reflectivity
measurements, we use variable slit conditions that roughly keep the
incident beam footprint constant on the liquid surface and also adjust
the detector slit accordingly. This increases the angular acceptance
of the detector with $Q_{z}$, which makes the dependence of $\sigma$
on $Q_{z}$ large enough to require a correction. To a first order
approximation, we correct for that by assuming $\sigma^{2}=a+bQ_{z}$,
where $a$ and $b$ are two fitting parameters. Other variable parameters
used to construct the ED across the interface include the thicknesses
of the slabs ($d_{i}$) and their corresponding electron densities
($\rho_{i}$) (for details on the definitions of parameters see Ref.\ \onlinecite{Bu2006}).
The minimum number of slabs is the one for which the addition of another
slab does not improve the quality of the fit, i.e. $\chi^{2}$, significantly.

To collect the fluorescence signal from the monolayer-counterion system,
we use an energy dispersive detector (X-PIPS Detector; model SXP8,
with a Multiport II multichannel analyzer both from Canberra Industries
Inc.). The pencil-like tip of the detector is lowered to the surface
in an aluminum well in front of a thin Kapton window located about
$\sim2$ cm above the liquid surface for fluorescence measurements
(the fluorescence-setup is under the same conditions as for reflectivity, namely,
the canister holding the trough is sealed and is purged with water
saturated helium). The major advantage of the fluorescence technique
is that it can distinguish between different
elements by their characteristic emission lines. In the present study
due to the low-yielded signal, the fluorescence signal is distinguished
and displayed over two regions corresponding to two regions of the
incident beam angles: below the critical angle for total reflection
($Q_{c}\sim0.022$ {\AA}$^{-1}$ with little variation due to salt
in the solutions), which is highly sensitive to elements at the surface,
and above the critical angle, which is sensitive to elements in the
bulk and surface of the solution. Below the critical angle, the evanescent
X-ray waves penetrate to $60-80$ {\AA} from the surface, which
makes the bulk contribution to the fluorescence intensity almost negligible
at concentrations smaller than $10^{-3}$ M.

\section{Experimental Results}
\noindent \begin{center}%
\begin{figure}[!]
\includegraphics[width=2.5in]{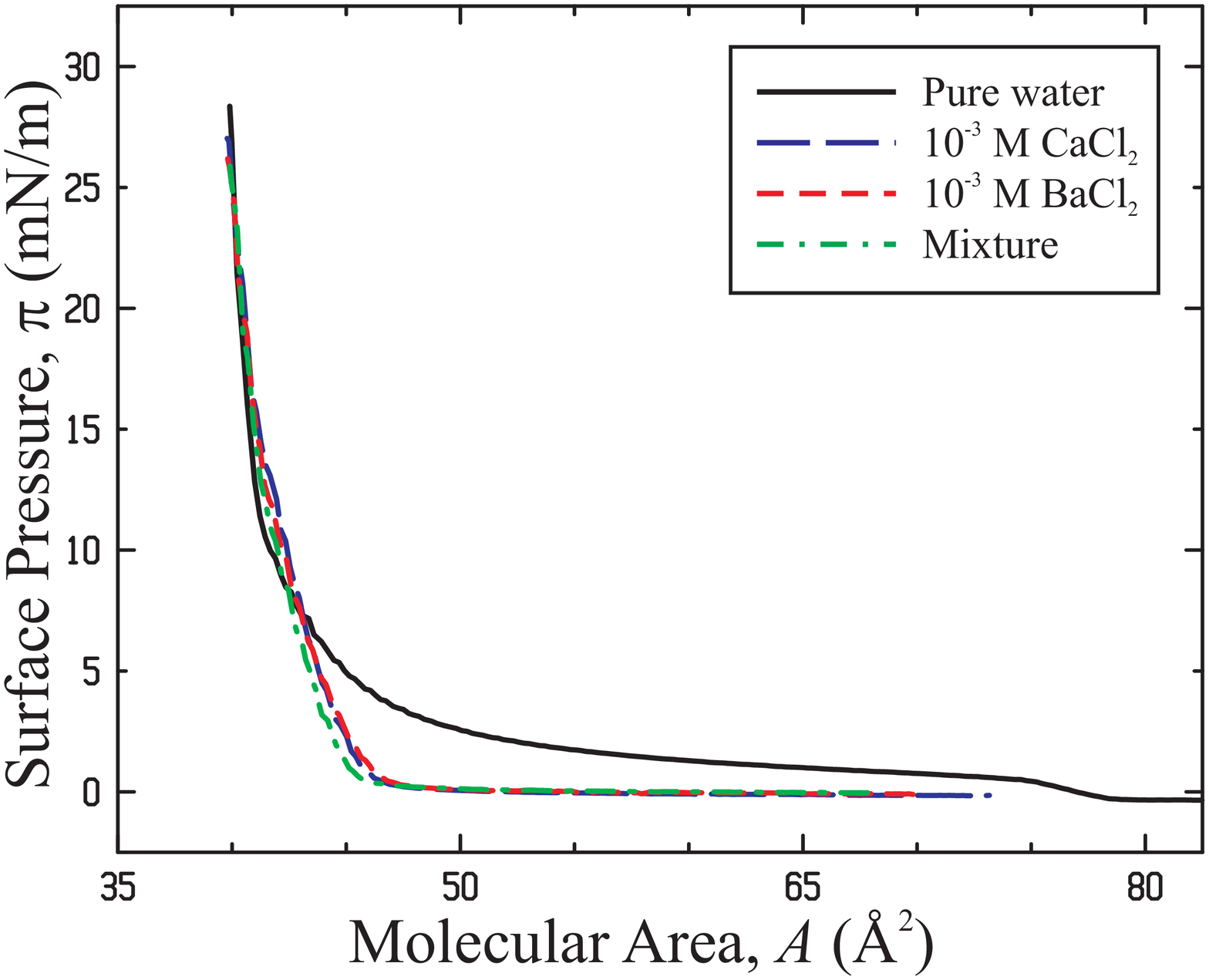}
\caption{\label{iso} Surface-pressure \textit{versus} molecular area ($\pi-A$)
isotherms of DMPA spread on water and dilute solutions, $10^{-3}$
M $\mathrm{CaCl_{2}}$, $\mathrm{BaCl_{2}}$, and a mixture solution
($5\times10^{-4}$M $\mathrm{CaCl_{2}}$ + $5\times10^{-4}$M $\mathrm{BaCl_{2}}$).
Although the isotherms on salts solution are different than that of
water they are hardly distinguishable with no specific features that
can be associated with the different ions.}
\end{figure}
\par\end{center}
\subsection{Isotherms}
Figure\ \ref{iso} shows surface pressure versus molecular area isotherms
for three different salt solutions (as indicated) at concentrations
of $10^{-3}$M with DMPA and pure water with DMPA. The isotherm of
DMPA on pure water lacks any distinct phase transitions with a smooth
increase with compression. For DMPA spread on salt solutions, two
slopes associated with crystalline tilted and untilted phases of the
hydrocarbon chains with respect to the surface normal can be clearly
identified. As the monolayer is compressed, it reaches the untilted
stage ($\pi\approx13$ mN/m) at the sufficiently low molecular area
of $\sim40$ {\AA}$^{2}$ per molecule. The isotherms of DMPA on
the different salt solutions vary just slightly, exhibiting little
or no specific features to the different ions in the solutions. All
fluorescence and reflectivity measurements were conducted in the untilted
crystalline stage ($\pi\sim25$ mN/m).

\noindent \begin{center}%
\begin{figure}[!]
\includegraphics[width=2.5in]{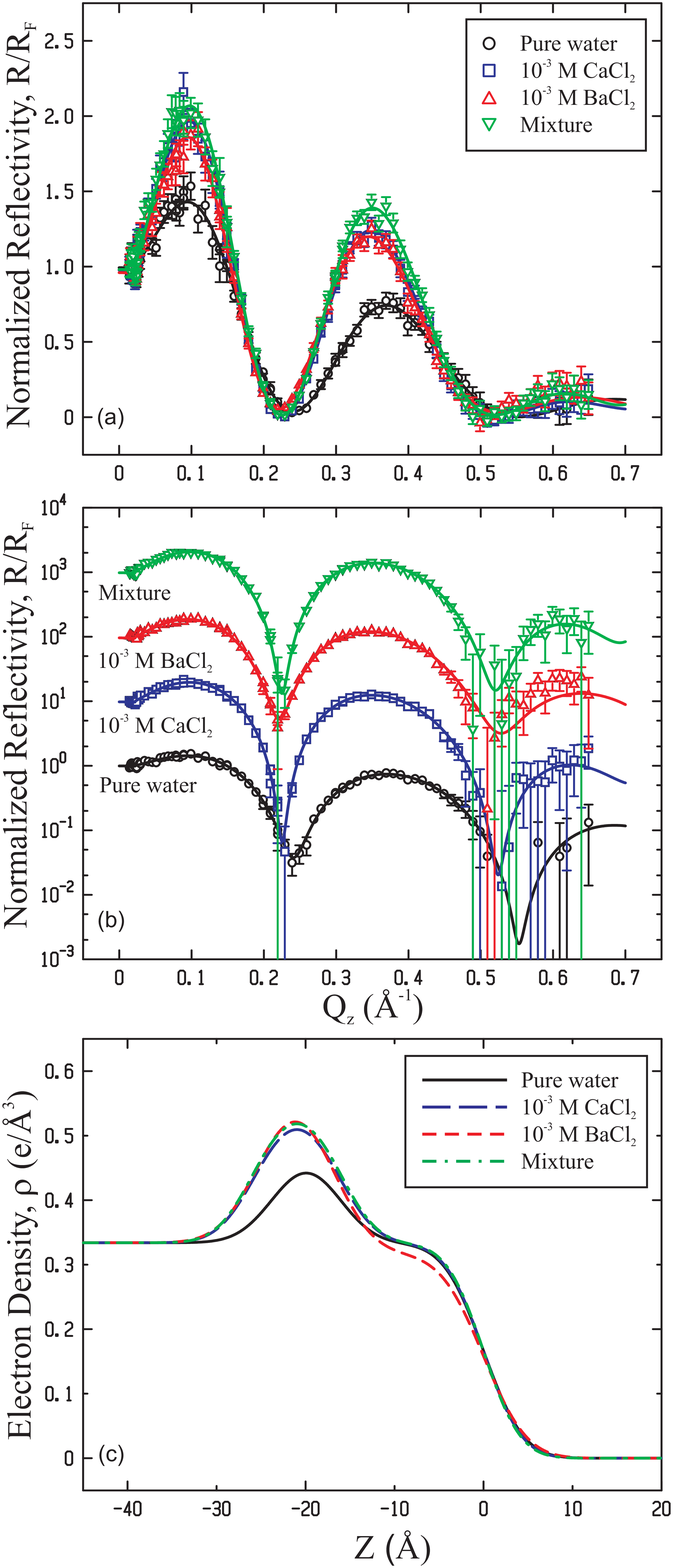}
\caption{\label{ref}(a) Normalized reflectivity data ($R/R_{F}$) \textit{versus}
momentum transfer $Q_{z}$ for each of the dilute solutions (as indicated)
and on water with DMPA on a linear scale. (b) The same data on a logarithmic
scale (curves and data are shifted by a decade each for clarity).
(c) Electron density profiles used to calculate the best fits shown
by solid lines in (a) and (b). The electron densities are generated
with the parameters listed in Table\ \ref{table}.}
\end{figure}
\par\end{center}

\subsection{X-ray Reflectivity}
Normalized reflectivity data ($R/R_{F}$, where $R_{F}$ is the calculated
reflectivity of pure water with an ideally flat interface; $\sigma=0$)
of DMPA monolayers spread on $10^{-3}$M solutions and on pure water
are shown in Fig.\ \ref{ref}. It is evident from Fig.\ \ref{ref}a
that the overall reflectivities for the samples prepared on salt solutions
are significantly higher than that of the pure water data. This is
a qualitative evidence that the ions ($\mathrm{Ba^{2+}}$ and/or $\mathrm{Ca^{2+}}$)
are accumulating at the surface. Also, there is a visible shift in
the minima of the reflectivities to a smaller $Q_{z}$ for the solutions
compared to water reflectivity, indicating a thicker layer due to
the association of the ions at the headgroup region.

The normalized reflectivity data were fit (solid lines Fig.\ \ref{ref}a
and b) to a two slab model, one associated with the hydrocarbon chains
of DMPA close to the vapor, and the other, contiguous to the bulk
solution, associated with the phosphatidic-acid headgroup and bound
ions and water molecules. The fitting parameters for the reflectivity
data are listed in Table\ \ref{table}. Using these parameters the
electron density profiles for each solution and for pure water are
obtained and plotted in Fig.\ \ref{ref}c. It is interesting to note
that the electron densities for divalent ion sub-phase are, within
experimental uncertainties, identical. Qualitatively, this surprising
result indicates there must be more calcium ions per DMPA than barium
ions as the two ions, $\mathrm{Ba^{2+}}$ and $\mathrm{Ca^{2+}}$,
have very different numbers of electrons, 18 and 54, respectively.
The phosphatidic-acid headgroup on pure water has a smaller thickness
and smaller electron density than those obtained for the salt solutions,
as expected since it does not have an electron-rich ion layer as with
the salt solutions. %
\begin{table}
\begin{centering}\begin{tabular}{ccccc}
\hline
{\scriptsize sample}&
{\scriptsize pure water}&
{\scriptsize $10^{-3}$M CaCl$_{2}$}&
{\scriptsize $10^{-3}$M BaCl$_{2}$}&
{\scriptsize mixture}\tabularnewline
\hline
\textbf{\scriptsize $d_{head}$}&
{\scriptsize $5.4_{-0.9}^{+5.8}$}&
{\scriptsize $9.2_{-3.4}^{+1.6}$}&
{\scriptsize $7.1_{-1.5}^{+4.9}$}&
{\scriptsize $9.9_{-4.1}^{+1.1}$}\tabularnewline
{\scriptsize $\rho_{head}$}&
{\scriptsize $0.53_{-0.1}^{+0.22}$}&
{\scriptsize $0.55_{-0.05}^{+0.1}$}&
{\scriptsize $0.63_{-0.17}^{+0.06}$}&
{\scriptsize $0.55_{-0.03}^{+0.12}$}\tabularnewline
\textbf{\scriptsize $d_{chain}$}&
{\scriptsize $17.2_{-3.2}^{+0.5}$}&
{\scriptsize $16.3_{-0.8}^{+1.7}$}&
{\scriptsize $17.5_{-2.5}^{+1.0}$}&
{\scriptsize $16.0_{-0.6}^{+2.0}$}\tabularnewline
{\scriptsize $\rho$$_{chain}$}&
{\scriptsize $0.33_{-0.03}^{+0.01}$}&
{\scriptsize $0.33_{-0.02}^{+0.01}$}&
{\scriptsize $0.32_{-0.02}^{+0.01}$}&
{\scriptsize $0.33_{-0.02}^{+0.01}$}\tabularnewline
{\scriptsize $\sigma$ at $0.3$ {\AA}$^{-1}$}&
{\scriptsize $3.6_{-0.8}^{+0.4}$}&
{\scriptsize $3.4_{-0.4}^{+0.3}$}&
{\scriptsize $3.8_{-1.1}^{+0.1}$}&
{\scriptsize $3.3_{-0.3}^{+0.4}$}\tabularnewline
{\scriptsize \# of ions}&
{\scriptsize N/A}&
{\scriptsize $2.2_{-0.6}^{+0.1}$}&
{\scriptsize $0.9_{-0.6}^{+0.1}$}&
{\scriptsize N/A}\tabularnewline
{\scriptsize \# of water}&
{\scriptsize $1.5_{-1.3}^{+9.4}$}&
{\scriptsize $6.8_{-5.6}^{+2.4}$}&
{\scriptsize $3.7_{-3.0}^{+6.4}$}&
{\scriptsize N/A}\tabularnewline
\end{tabular}\par\end{centering}
\caption{\label{table} Parameters used to fit the reflectivity data as well
as the derived number of ions and water molecules at the interface.
Uncertainties are obtained by fixing a parameter at values away from
the optimum and readjusting all of the other parameters to a new minimum
until $\chi^{2}$ increases by 25\%, and by imposing limiting values
on some of the parameters to keep them within physical range; for
instance, the electron density of the hydrocarbon chains was kept
within 0.32 $\pm0.03$.}
\end{table}

From the structural parameters in Table\ \ref{table}, one can derive
the number of ions per DMPA molecule at the surface by applying volume
constraints and assuming the co-ions (e.g., $\mathrm{Cl^{-}}$) do
not penetrate into the headgroup region (as shown below). We use the
following equations to solve for the number of water molecules ($n_{w}$)
and the number of ions ($n_{i}$) per phosphatidic acid in the headgroup
slab.\cite{Gregory1997} \begin{equation}
Ad_{h}=V_{h}+n_{w}V_{w}+n_{i}V_{i}\label{constraints0}\end{equation}
 \begin{equation}
Ad_{h}\mathnormal{\mathrm{\rho}}_{h}=N_{h}+n_{w}N_{w}+n_{i}N_{i},\label{constraints}\end{equation}
 where $A$ is the molecular area obtained from the isotherm and diffraction
data,\cite{Vaknin1991,Gregory1997,Vaknin2003} $d_{h}$ and $\rho_{h}$
are the thickness and electron density of the head group, respectively,
$N_{i}$ is the number of electrons per molecule (for water molecule
$N_{w}=10$, and for $\mathrm{Ca^{2+}}$ and $\mathrm{Ba^{2+}}$ 18
and 54 electrons, respectively). The molecular area can be confirmed
self consistently by the electron density of the hydrocarbon chains
$\rho_{chain}$ from the relation $A=N_{chain}/(d_{chain}\rho_{chain})$.
The volume of the phosphatidic headgroup, $V_{h}$, is obtained by
the relations above, eqs.\ \ref{constraints0} and \ref{constraints},
using the the electron density profile parameters for DMPA spread
on pure water. The precise position that parses the molecule into
the hydrocarbon chain compartment and the headgroup region influences
just slightly the derived parameters, and this change is accounted
for in the uncertainties of the parameters given in Table\ \ref{table}.
As seen in the table and as expected by the electron-density profiles,
the number of $\mathrm{Ca^{2+}}$ and the number of $\mathrm{Ba^{2+}}$
ions per DMPA is not the same but rather seems to favor $\mathrm{Ca^{2+}}$
by a $\sim2:1$ ratio for the unmixed salt solutions.

\noindent \begin{center}%
\begin{figure}[!]
\includegraphics[width=2.5in]{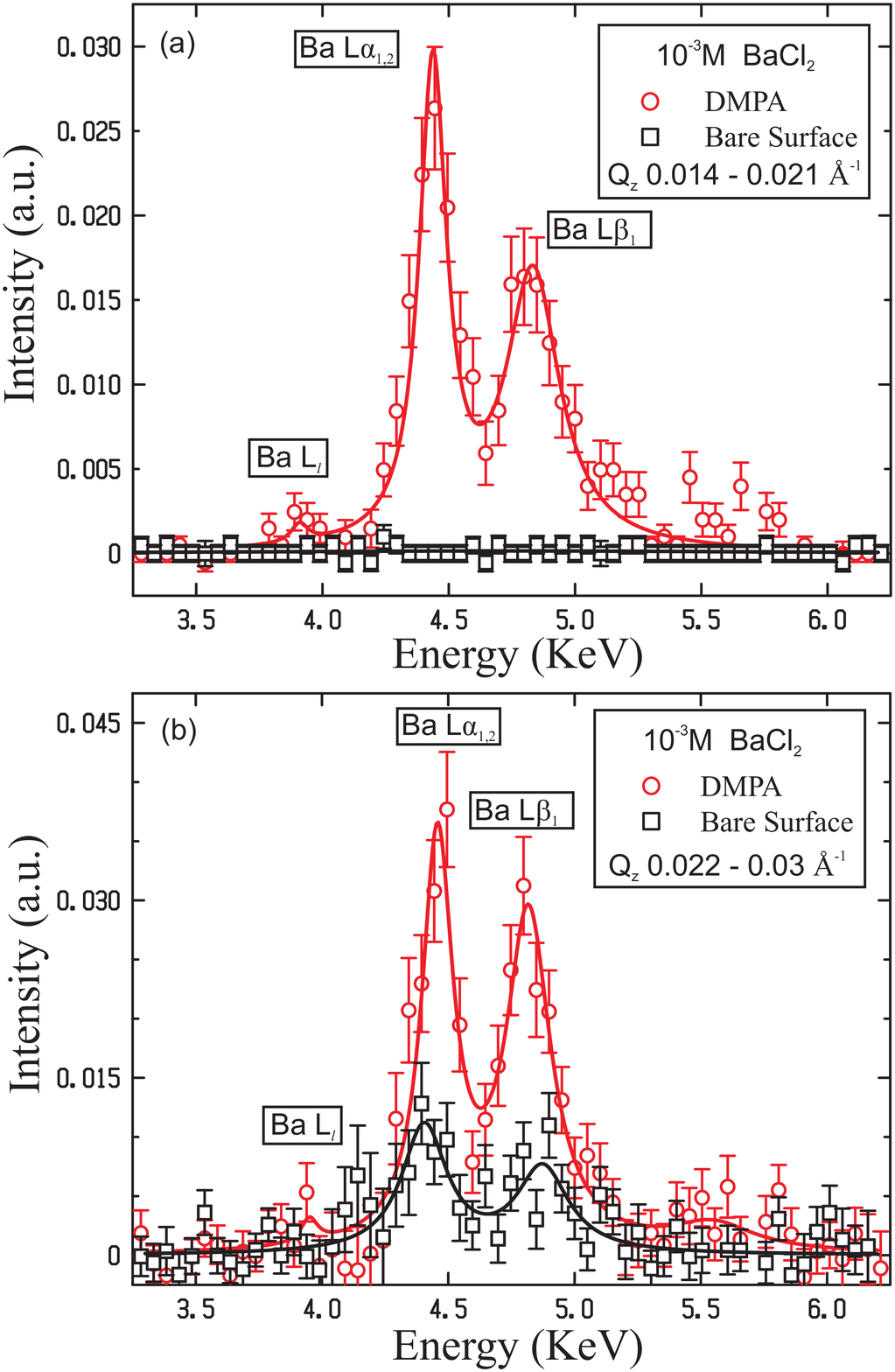}
\caption{\label{ba1lcut}(a) Fluorescence signals below the critical angle
for $10^{-3}$M BaCl$_{2}$ of the solution only and the same solution
with the DMPA monolayer as indicated. (b) Fluorescence data above
the critical angle from the same two samples. The signals above and
below the critical angle were integrated over a $Q_{z}$ range as
indicated in the figures.}
\end{figure}
\par\end{center}

\subsection{Fluorescence}
Fluorescence scans were performed on all the solutions with and without
DMPA, and the background data was obtained from a pure water sample
without DMPA and subtracted from all data to remove electronic and
stray signals from the trough. The data from the fluorescence were
then integrated over $Q_{z}$ ranges to improve signal to noise ratio
as described above. Figure \ \ref{ba1lcut} shows fluorescence data
below (a) and above (b) the critical angle for $10^{-3}$M $\mathrm{BaCl_{2}}$
with and without DMPA, respectively. The intensity was integrated
over the range shown in the figures. As shown in Fig.\ \ref{ba1lcut}(a)
(below the critical angle), without a monolayer the evanescent X-ray
beam does not penetrate far enough into the bulk to generate any visible
fluorescence intensity. However, with the DMPA monolayer on, the $\mathrm{L}\alpha$
(4.46 keV) and $\mathrm{L}\beta_{1}$ (4.8 keV) emission lines for
$\mathrm{Ba^{2+}}$ are clearly visible, which indicates that the
concentration of ions at the interface is much larger compared to
that of the bulk concentration. In Fig.\ \ref{ba1lcut}(b), emission
lines can be seen in both the solution without and with DMPA. However,
the solution with the monolayer has an enhancement for each emission
line as the ions at the surface contribute significantly to the overall
intensity. The difference in the intensities of the two data sets
in Fig.\ \ref{ba1lcut}(b) quantifies the contribution of the ions
at the surface. Spectra of $\mathrm{K}\alpha\approx3.7$ keV and $\mathrm{K}\beta\approx4.03$
keV emission lines for the $\mathrm{CaCl_{2}}$ solution (not shown)
essentially show similar trends. Figure \ \ref{conclut} shows
that the emission lines below the critical angle (contributed primarily
by surface) are practically identical at $10^{-3}$M and $10^{-2}$M
$\mathrm{BaCl_{2}}$, indicating that the monolayer is saturated with
ions for concentrations stronger than $10^{-3}$M. $\mathrm{CaCl_{2}}$
and the mixture solutions display similar binding phenomena at these
concentrations.

\noindent \begin{center}%
\begin{figure}[!]
\includegraphics[width=2.5in]{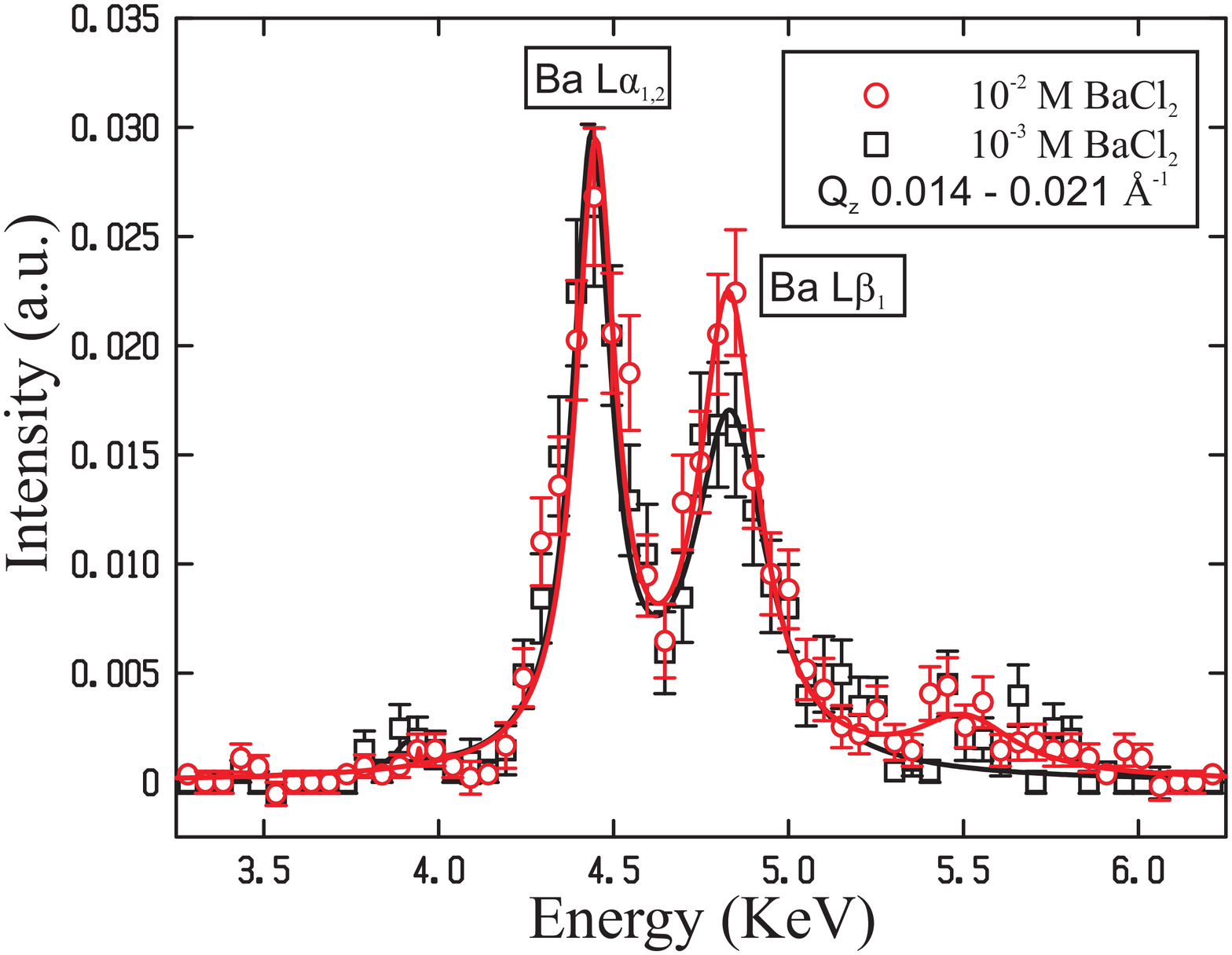}
\caption{\label{conclut} Integrated fluorescence signals below the critical
angle for solutions of $10^{-2}$ M and $10^{-3}$ M $\mathrm{BaCl_{2}}$
showing the binding of ions reaches saturation for concentrations
$\lesssim$ $10^{-3}$ M.}
\end{figure}
\par\end{center}
To get more quantitative results we define a ''standard ratio''
as the ratio of fluorescence signals for specific emission lines from
two different elements in equal quantities in the bulk solution (e.g.,
$\mathrm{\mathrm{Ca^{2+}}/Ba^{2+}}$). From Fig.\ \ref{baseline}(a),
pure solutions without DMPA, we establish a \char`\"{}standard ratio\char`\"{}
(by comparing the peak intensities of $\mathrm{Ba^{2+}}$ $\mathrm{L\alpha}$
and $\mathrm{Ca^{2+}}$ $\mathrm{K\alpha}$) of the emission lines
generated by equal concentrations of $\mathrm{Ba^{2+}}$ and $\mathrm{Ca^{2+}}$
ions at $\sim6:1$ in favor of $\mathrm{Ba^{2+}}$. Comparing the ratio between the same peak intensities provides the relative population of each ion at the interface. From Fig. \ \ref{baseline}(b) we find
the $\mathrm{Ba^{2+}}$/$\mathrm{Ca^{2+}}$ ratio of the pure solutions
reduces to $\sim 3:1$ and for the mixed solution the ratio is even
lower $\sim1.5:1$. This shows that $\mathrm{Ca^{2+}}$ in unmixed
solution populates the PA interface twice as much as the $\mathrm{Ba^{2+}}$
does, consistent with X-ray reflectivity results.  In the Ba/Ca 1:1 mixed
solutions calcium ions are more likely to associate with the PA at the
interface by a ratio of $\sim 4:1$ over $\mathrm{Ba^{2+}}$.
\noindent \begin{center}%
\begin{figure}[!]
\includegraphics[width=2.5in]{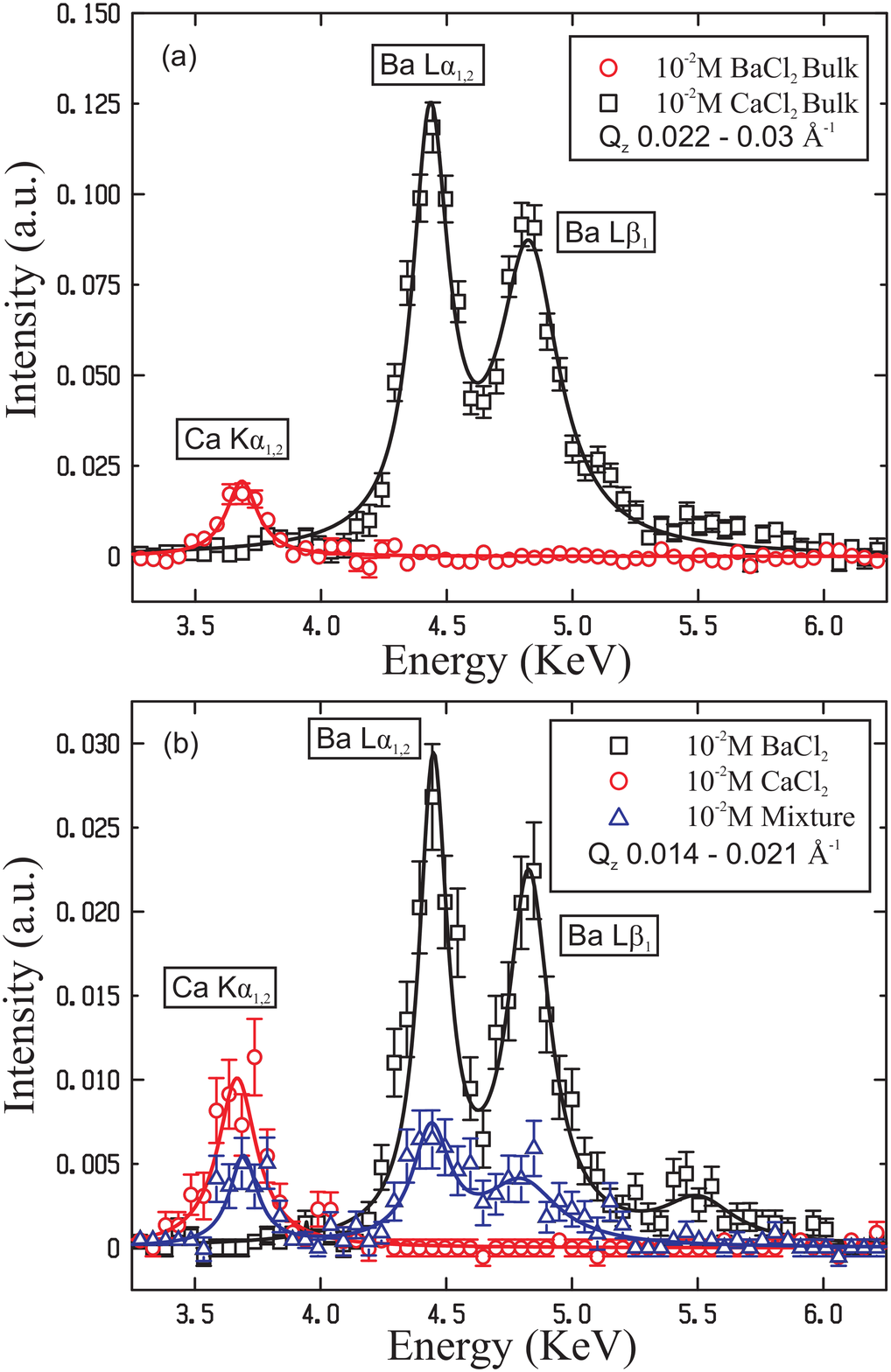}
\caption{\label{baseline}(a) Integrated fluorescence signals above the critical
angle showing the relative intensities of Ba and Ca emission lines
from the solutions only at $10^{-2}$ M (no monolayer). This data
is used to calibrate the relative intensities from the same number
of Ba and Ca ions. (b) Integrated fluorescence signals below the critical
angle for DMPA monolayers spread on $\mathrm{BaCl_{2}}$, $\mathrm{CaCl_{2}}$
and 1:1 mixture at a nominal $10^{-2}$ M. This indicates the amount
of bound $\mathrm{Ba^{2+}}$ to PA is significantly smaller than that
of $\mathrm{Ca^{2+}}$ for DMPA on the mixture.}
\end{figure}
\par\end{center}

\section{Discussion and Conclusions}%
\begin{figure}[!]
\includegraphics[width=2.5in]{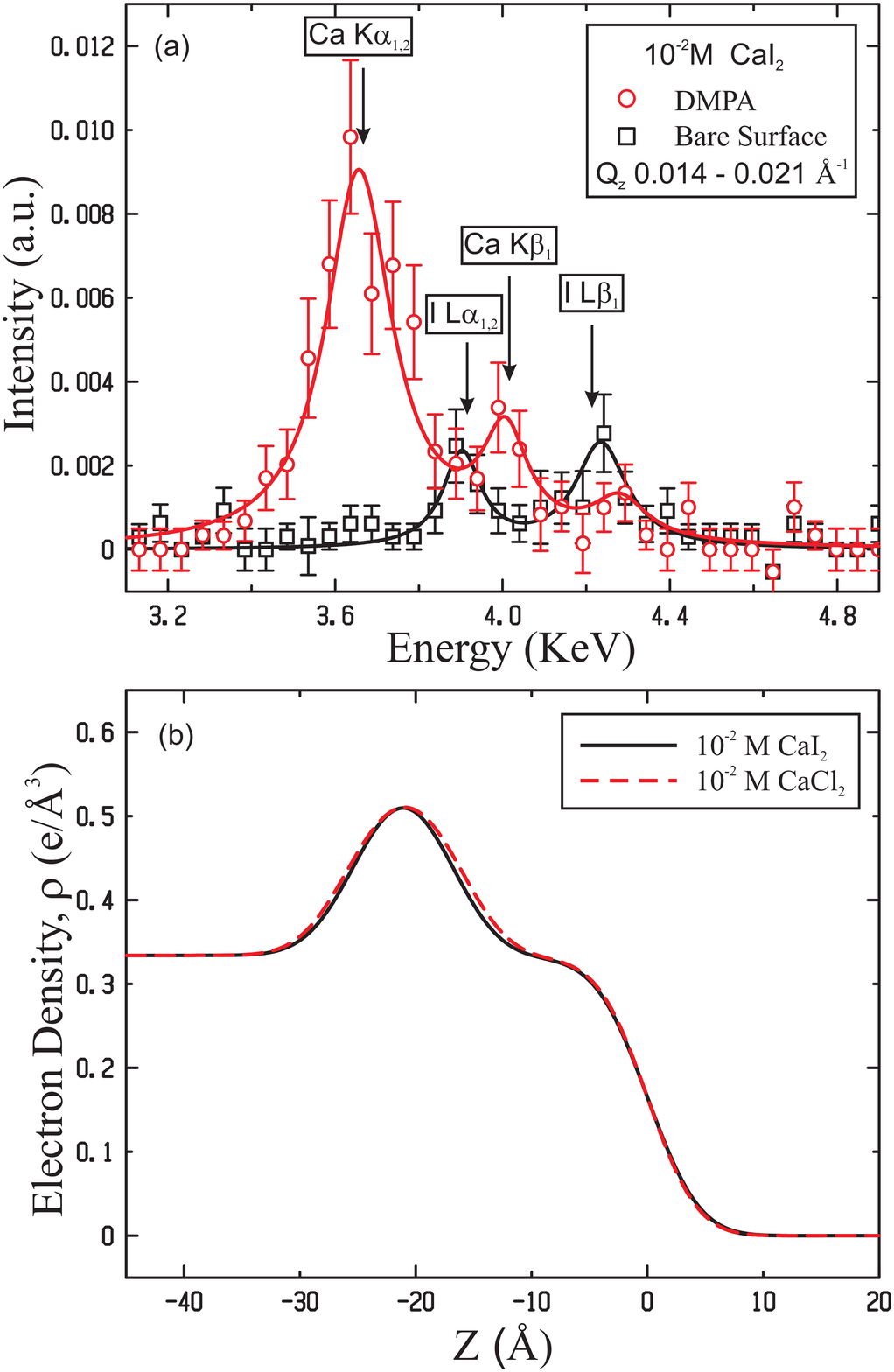}
\caption{\label{iodine}(a) Fluorescence signals from the surface of $\mathrm{CaI_{2}}$
($10^{-2}$ M) solution only, and the same solution with the DMPA
monolayer on. At this relatively high concentration the signal from
iodine is clearly seen before spreading the monolayer whereas the
signal form calcium is very weak to detect. With the DMPA monolayer
on, the Ca signal is enhanced whereas the signal from the iodine slightly
weakens. This shows there is no enrichment of the iodine co-ion at
the interface due to possible charge-inversion. (b) The electron density
profiles for films of DMPA spread on $10^{-2}$ M $\mathrm{CaI_{2}}$
and $10^{-2}$ M $\mathrm{CaCl_{2}}$ solutions. This shows the amount
of $\mathrm{I}^{-}$ is negligible at the interface although the number
of $\mathrm{Ca^{2+}}$ ions exceeds those necessary to neutralize
the phosphatidic-acid.}
\end{figure}

The objective of this study has been to explore the binding specificity
of divalent ions in solutions to a phosphatidic acid charged surface
and in particular, to determine which of the two $\mathrm{Ba^{2+}}$
or $\mathrm{Ca^{2+}}$ is more favorable to binding to the DMPA monolayer.
Our results show that, in a 1:1 mixed solution, $\mathrm{Ca^{2+}}$
is favored over $\mathrm{Ba^{2+}}$ by almost a 4:1 ratio. This is
strikingly different from the results of Shapovalov and coworkers\cite{Shapovalov2007}
who found that $\mathrm{Ba^{2+}}$ outnumbers $\mathrm{Ca^{2+}}$
by a 10:1 in binding to a behenylsulfate monolayer. Clearly, the only
possible origin of such a discrepancy is a strong specificity of ionic
and headgroup interactions, and cannot be quantitatively accounted
for by theories based on classical electrostatic interactions, even
if the ionic size is properly accounted for. Furthermore, our finding
that $\mathrm{Ca^{2+}}$ binds more strongly than $\mathrm{Ba^{2+}}$
can only be understood if the ion dehydrates upon binding. Quantitative
theoretical account for these effects remains an outstanding theoretical
problem.

Our results also imply charge inversion at the surface in particular,
for the $\mathrm{CaCl_{2}}$ and $\mathrm{CaI_{2}}$ solutions, as
shown below. This happens when the charge density of the counterions
exceeds the surface charge density exerted by the phosphatidic monolayer.
Our analysis shows there there is one $\mathrm{Ba^{2+}}$ per DMPA molecule,
which is just enough to neutralize the surface charges (doubly charged
phosphatidic acid), whereas there is more than one $\mathrm{Ca^{2+}}$
ion per DMPA molecule at the surface. This suggests charge inversion,
from negatively charged to positively charged interface. In the fluorescence
experiments above, we did not detect strong emission signals from
$\mathrm{Cl^{-}}$ to suggest that there exists an accumulation of
the co-ion (Cl$^{-}$) at the interface due to charge inversion. To
further explore if the anions are rising to the charge-inverted surface,
we performed another experiment using $10^{-2}$ M $\mathrm{CaI_{2}}$
solution with DMPA as a monolayer to improve the fluorescent signal
from the co-ion, in this case from $\mathrm{I}^{-}$. Figure\ \ref{iodine}(a)
shows spectra below the critical angle, indicating surface enrichment
of $\mathrm{Ca^{2+}}$ but no significant difference in the emission
lines of $\mathrm{I^{-}}$ before and after the spreading of DMPA.
We have also measured the reflectivity from the same DMPA monolayer
on $\mathrm{CaI_{2}}$ solution to detect any changes in the electron
density near the headgroup region compared to that of DMPA on $\mathrm{CaCl_{2}}$
solution, that may indicate the accumulation of $\mathrm{I^{-}}$
at the interface. Figure\ \ref{iodine}(b) shows the ED profiles
of the films on both solutions ($\mathrm{CaCl_{2}}$ or $\mathrm{CaI_{2}}$)
are practically the same. We therefore conclude that if there are
co-ions (iodine or chlorine ions) at the surface due to charge inversion, their concentration
is below 0.1 iodine ion per PA molecule according to our detection
sensitivity, or that the $\mathrm{Ca^{2+}}$ ions migrate to the surface
partially as $\mathrm{CaOH^{+}}$.

To summarize, in numerous physiological studies such as those of the 'Ca receptor'
in preganglionic nerve terminals (interacting primarily with acetyl-choline)
it has found that $\mathrm{Ba^{2+}}$ ions, do not appear to compete
with $\mathrm{Ca^{2+}}$ for the receptor\cite{McLachlan1977}. This is consistent
with our studies that show stronger affinity of Ca over Ba to negatively charged PA interface.
Systematic studies of model systems using spectroscopic X-ray techniques that are ion specific open the
door to unraveling the mechanism of this and related physiological
phenomena.

\begin{acknowledgments}
We wish to thank Alex Travesset for helpful discussions during the
course of this work, and for his comments and suggestions on the manuscript.
This manuscript has been authored, in whole or in part, under
Contract No. DE-AC02-07CH11358 with the U.S. Department of Energy.
\end{acknowledgments}


\begin{references}

\bibitem{Meijer2003} Meijer, H. G. H.; Munnik, T. \textit{Annu. Rev.  Plant Biol.} \textbf{2003}, \textit{54}, 265.
\bibitem{Ishii2004} Ishii, I.; Fukushima, X. Ye.; Chun, J. \textit{Annu. Rev. Biochem.} \textbf{2004}, \textit{73}, 321.
\bibitem{Wang2006} Wang, X.; Devaiah, S. P.; Zhang, W.; Welti, R. \textit{Prog. Lipid Res.} \textbf{2006}, \textit{45}, 250.
\bibitem{Kooijman2007} Kooijman, E. E.; Tielman, D. P.; Testerink, C.; Munnik, T.; Rijkers, D. T. S.; Burger, K. N. J.; de Kruiiff, B.  \textit{J. Biol. Chem.} \textbf{2007}, \textit{282}, 11356.

\bibitem{Faraudo2007b} Faraudo, J,; Travesset, A. \textit{Biophys.
J.} \textbf{2007}, \textit{92}, 2806; and \textit{Coll. Surf. A} \textbf{2007},
\textit{300}, 287.

\bibitem{Vaknin2003} Vaknin, D.; Kruger, P.; Losche, M. \textit{Phys.
Rev. Lett}. \textbf{2003}, \textit{90}, 178102/1-4.

\bibitem{Pittler2006} Pittler, J.; Bu, W.; Vaknin, D.; Travesset,
A.; McGillivray, D. J.; Losche, M. \textit{Phys. Rev. Lett.} \textbf{2006},
\textit{97}, 046102/1-4.

\bibitem{Bloch1985} Bloch, J. M.; Sansone, M.; Rondelez, F.; Peiffer,
D. G.; Pincus, P.; Kim, M. W.; Eisenberger, P. M. \textit{Phys. Rev.
Lett.} \textbf{1985}, \textit{54}, 1039.

\bibitem{Bloch1988} Bloch, J.M.; Yun, W.B.; Yang, X.; Ramanathan,
M.; Montano, P. A.; Capasso, C. \textit{Phys. Rev. Lett.} \textbf{1988},
\textit{61}, 2941.

\bibitem{Kjaer1989} Kjaer, K.; Als-Nielsen, J.; Helm, C.; Tippman-Krayer,
P.; Mohwald, H. \textit{J. Phys. Chem.} \textbf{1989}, \textit{93},
3200.

\bibitem{Jun1990} Jun, J. W.; Bloch, J. M. \textit{J. Appl. Phys.}
\textbf{1990}, \textit{68}, 1421.

\bibitem{Daillant1991} Daillant, J.; Bosio, L.; Benattar, J. J.;
Blot, C. \textit{Langmuir} \textbf{1991}, \textit{7}, 611.

\bibitem{Jacquemain1991} Jacquemain, D.; Wolf, S. G.; Leveiller,
F.; Deutsch, M.; Kjaer, K.; Als-Nielsen, J.; Lahav, M.; Leiserowitz,
L. \emph{Angew. Chem.} \textbf{1992}, \emph{31}, 130.

\bibitem{Novikova2003} Novikova, N. N.; Zheludeva, S. I.; Konovalova,
O. V.; Kovalchuk, M. V.; Stepina, N. D.; Myagkov, I. V.; Godovsky,
Y. K.; Makarova, N. N.; Tereschenko, E. Y; Yanusova, L. G. \textit{J.
Appl. Crystallogr}. \textbf{2003}, \textit{36}, 727.

\bibitem{Zheludeva2003} Zheludeva, S. I.; Novikova, N. N.; Konovalov,
O. V.; Kovalchuk, M. V.; Stepina, N. D.; Tereschenko, E. Y. \textit{Mater.
Sci. Eng. C} \textbf{2003}, \textit{23}, 567.


\bibitem{Bu2005} Bu, W.; Vaknin, D.; Travesset, A. \textit{Phys.
Rev. E} \textbf{2005}, \textit{72}, 060501(R)/1-4.

\bibitem{Bu2006} Bu, W.; Vaknin, D.; Travesset, A. \textit{Langmuir}
\textbf{2006}, \textit{22}, 5673.

\bibitem{Bu2006b} Bu, W.; Ryan, J. P.; Vaknin, D. \textit{J. Synchrotron
Rad.} \textbf{2006}, \textit{13}, 459.

\bibitem{Shapovalov2006} Shapovalov, V. L.; Brezesinski, G. \textit{J.
Phys. Chem. B} \textbf{2006}, \textit{110}, 10032.

\bibitem{Shapovalov2007} Shapovalov, V. L.; Ryskin, M. E.; Konovalov,
O. V.; Hermelink, A.; Brezesinski, G. \textit{J. Phys. Chem.} \textbf{2007},
\textit{111}, 3927.

\bibitem{Ninham1971} Ninham, B W.; Parsegian, V. A.; \textit{J. Theor.
Biol.} \textbf{1971}, \textit{31}, 405.

\bibitem{Bloch1990} Bloch, J. M.; Yun, W. \textit{Phys. Rev. A} \textbf{1990},
\textit{41}, 844.

\bibitem{Israelachvili2000} Israelachvili, J. \emph{Intermolecular
and surface forces}, Academic Press, London, 2000.

\bibitem{Andelman1995} Andelman, D. \textit{Handbook of Biological
Physics}, vol. I, R. Lipowsky.; E. Sackmann (Eds.), \textit{Electrostatic
Properties of Membranes: The Poisson-Boltzmann Theory}, p 603-641
Chapter 12 (Elsevier Science, Amsterdam, 1995).

\bibitem{Luo2006} Luo, G. M.; Malkova, S.; Yoon, J.; Schultz, D.
G.; Lin, B. H; Meron, M.; Benjamin, I.; Vanysek, P.; Schlossman, M.
L. \textit{Science} \textbf{2006}, \textit{311}, 216.

\bibitem{Giewekemeyer2007} Giewekemeyer K.; Salditt, T. \textit{Euro.
Phys. Lett.} \textbf{2007}, \textit{79}, 18003/1-18003/6.

\bibitem{Koelsch2007} Koelsch, P.; Viswanath, P.; Motschmann, H.;
Shapovalov, V. L.; Brezesinski, G.; Möhwald, H.; Horinek, Dominik;
Netz, Roland R.; Giewekemeyer, K.; Alditt, T. S.; Schollmeyer, H.;
von Klitzing, Regine.; Daillant, J.; Guenoun, P. \textit{Coll. Surf.
A} \textbf{2007}, \textit{303}, 110.

\bibitem{Travesset2006} Travesset, A.; Vaknin, D. \textit{Europhys.
Lett}. \textbf{2006}, \textit{74}, 181.

\bibitem{Faraudo2007a} Faraudo, J.; Travesset, A. \textit{J. Phys.
Chem. C} \textbf{2007}, \textit{111}, 987.

\bibitem{Biesheuvel2007}Biesheuvel, P. M.; van Soestbergen, M. \textit{J.
Coll. and Inter. Sci.} \textbf{2007}, \textit{316}, 490.

\bibitem{Vaknin1991}Vaknin, D.; Kjaer, K.; Als-Nielsen, J.; Losche,
M. \textit{Biophys. J.} \textbf{1991}, \textit{59}, 1325.

\bibitem{Kjaer1994} Kjaer, K. \textit{Physica B} \textbf{1994}, \textit{198},
100.

\bibitem{Vaknin2003b} Vaknin, D. \textit{Characterization of Materials},
Kaufmann, E. N., Ed,; Wiley: New York, 2003; Vol. 2, p 1027.

\bibitem{Parrat1954}Parratt, L. G. \textit{Phys. Rev.} \textbf{1954},
\textit{59}, 359.

\bibitem{Braslau1988}Braslau, A.; Pershan, P. S.; Swislow, G.; Ocko,
B. M.; Als-Nielsen, J. \textit{Phys. Rev. A} \textbf{1988}, \textit{38},
2457.

\bibitem{Ocko1994}Ocko, B. M.; Wu, X. Z.; Sirota, E. B.; Sinha, S.
K.; Deutsch, M. \textit{Phys. Rev. Lett.} \textbf{1994}, \textit{72},
242.

\bibitem{Gregory1997} Gregory, B. W.; Vaknin, D.; Gray, J. D.; Ocko,
B. M.; Stroeve, P.; Cotton, T. M.; Struve, W. S. \emph{J. Phys. Chem.
B} \textbf{1997}, \emph{101}, 2006.

\bibitem{McLachlan1977} E. M. McLachlan, E. B.; \textit{J. Physiol.}
\textbf{1977}, \textit{267}, 497.\end{references}
\end{document}